\documentclass[english,aps, superscriptaddress, floatfix, 12pt]{revtex4}
\usepackage[T1]{fontenc}
\usepackage[latin1]{inputenc}
\usepackage{graphicx}
\usepackage{axodraw}
\usepackage{epsfig}

\makeatletter

\usepackage{babel}
\makeatother
\begin{document}

\title{A Heavy Quark-Antiquark Pair in Hot QCD}

\author{D. Antonov, S. Domdey, H.-J. Pirner\\
{\it Institut f\"ur Theoretische Physik, Universit\"at Heidelberg,\\
Philosophenweg 19, D-69120 Heidelberg, Germany}}

\begin{abstract}
  Thermodynamics of a heavy quark-antiquark pair in SU(3)-QCD is studied
  both below and above the deconfinement critical temperature $T_c$.
  In the quenched case, a model of the string passing through heavy
  valence gluons yields a correct estimate of $T_c$ and the critical behavior of the string tension below $T_c$.
  For two light flavors, entropy and internal energy below $T_c$ can be obtained from the partition function of heavy-light 
  mesons and baryons. To calculate the free energy of the system above $T_c$, 
  we apply second-order perturbation theory in the interaction 
 of the quark-gluon plasma constituents with the static quark-antiquark pair. 
The results for the entropy and internal energy, obtained both below 
 and above $T_c$, are compared with recent lattice data. 

\end{abstract}
\maketitle
\section{Introduction}

Recent RHIC experiments \cite{Adcox:2004mh,Arsene:2004fa,Back:2004je,Adams:2005dq}
suggest that the quark-gluon plasma may be more like a perfect
liquid \cite{revs}, where the mean free path of a particle is much smaller than
the inter-particle distance.  Values of the mean free path following
from {\it perturbative} calculations~\cite{ay}, however, are {\it
large}. This makes perturbative approaches to the plasma questionable.

Lattice simulations of the quark-gluon plasma also need
non-perturbative methods for their explanation. For instance, it has
recently been argued~\cite{yuasim} that non-perturbative chromo-{\it
  electric} fields can describe the singlet free energy of a static
quark-antiquark pair at large distances which remains quite sizeable
above the deconfinement temperature $T_c$.  Other lattice data calling
for a theoretical explanation are the anomalously large maxima of the
internal energy and entropy of the static quark-antiquark pair in
unquenched QCD around $T_c$~\cite{latint1} (see Ref.~\cite{lat} for a review). Equally surprising is the
rapid fall-off of these quantities right after the phase transition.
This paper aims to analyze what kind of nonperturbative
physics is necessary to understand these lattice simulations. We consider three cases:
In finite-temperature quenched QCD  the $Q\bar Q$ system
contains many valence
gluons~\cite{L, spect}, which may form a gluon chain.  We demonstrate
that the high entropy generated by this object gives a reasonable prediction
for the critical behavior of the string tension near $T_c$.
In unquenched QCD  for $T<T_c$  the QCD string may break and the produced 
light quark and antiquark interact with the heavy  $Q\bar Q$ pair to
form mesons. In unquenched QCD for  $T>T_c$ the constituents of the quark-gluon plasma 
interact via screened gluon exchange with the static quark and antiquark. These three 
simple models are supposed to catch the main physical aspects of
the heavy-quark system in hot QCD. They cannot represent a unified
picture due to their inherent approximations. By comparison with
the lattice simulations one sees how realistic the so obtained picture
is.

The paper is organized as follows. In the next section, we consider
quenched SU($N_c$) QCD below $T_c$, calculate the partition function of
a gluon chain, and make predictions for $T_c$ and the effective string
tension $\sigma(T)$. In section~III, using $\sigma(T)$, we calculate
within the relativistic quark model the partition function of
heavy-light mesons and baryons. The entropy and internal energy
stemming from this partition function are compared with the
corresponding lattice data.  In section~IV, we calculate the 
interaction energy of quarks, antiquarks and gluons with 
the heavy quark-antiquark pair immersed in the plasma. We impose that the plasma
particle-number densities vanish at $T=T_c$. With this constraint
implemented we calculate the change of the  entropy and internal energy in second
order perturbation theory due to the  quark and antiquark pair.
Finally, we estimate contributions coming from possible light $q\bar
q$ octet bound states, which may exist after deconfinement. In
section~V, we summarize the main results of the paper.

\section{Static $Q\bar Q$-pair with gluons at $T<T_c$}

The $Q\bar Q$-string which sweeps out the flat surface
of the corresponding Wilson loop is produced by soft stochastic
gluonic fields. In addition, fluctuations of the gauge field exist,
which lead to string vibrations. These fluctuations may be related to
valence gluons through which the $Q\bar Q$-string passes. 
At asymptotically large $Q\bar
Q$-separations considered here the string passes through many valence
gluons and forms a gluon chain~\cite{Greensite}. Further,
the energy of one string bit between two nearest gluons in the chain
is constant. As long as the thermal mass of a valence gluon is smaller than this energy, 
gluons move together with the string and do not affect the global dynamics of the string (see Fig.~1). 
However, at a certain temperature $T_0$ the gluon's thermal mass $(\propto T)$
becomes larger than the energy of one string bit. From this
temperature on, the system looks totally different, since gluons from
the string's standpoint are now nearly static.  Therefore at
$T_0<T<T_c$ a gluon chain becomes a sequence of static nodes with
adjoint charges, connected by independently fluctuating string bits
(see Fig.~2). At the moment of formation of such a chain, its
end-point originating from the heavy $Q$ performs a {\it random walk}
towards $\bar Q$ over the lattice of static nodes.  The entropy of
such a random walk turns out to be large, namely proportional to its
length, and eventually leads to the deconfinement phase transition in
this model. The reason for a large entropy is that color may alter
from one node of the random walk to another, i.e. every string bit may
transport each of the $N_c$ colors. A similar picture has been
proposed by Greensite and Thorn~\cite{Greensite} as a time cut of a large Wilson
loop leading to the above sequence of valence gluons arising from
double lines of neighboring plaquettes.  Therefore, the total number
of states of the gluon chain grows exponentially with its length, $L$,
as $N_c^{L/a}$, where $a$ is the length of one bit of the chain.

\begin{figure}
\includegraphics[clip,scale=0.5]{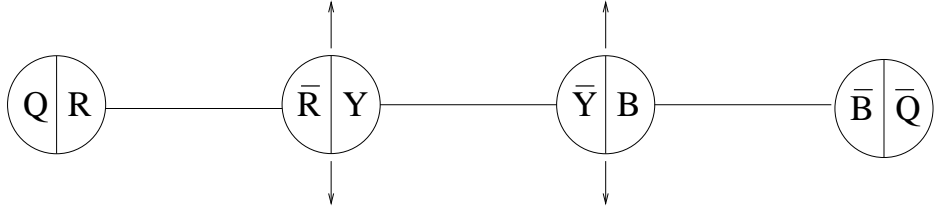}
\caption{Gluon chain for $T<T_0$ where valence gluons move together with the string.}
\end{figure}
\begin{figure}
\includegraphics[clip,scale=0.5]{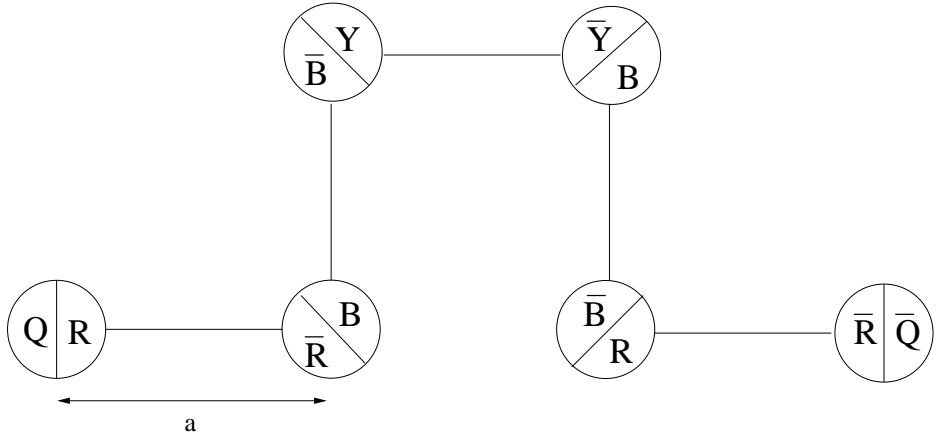}
\caption{Gluon chain for $T>T_0$ where the valence gluons become static, but color may change from one string bit to another.} 
\end{figure}

To quantify these considerations, let us begin with considering a free random walk, which is performed
over a $d$-dimensional Euclidean hypercubic lattice (in our case $d=4$) with the spacing
$a$.  At zero temperature, a walker, starting
from the origin, arrives at a distance $R$ ($R\equiv|{\cal R}|$,
where ${\cal R}$ is a $d$-dimensional vector) with the probability~\cite{idr}
\begin{equation}
\label{1}
P(s,R)=\frac{1}{(4\pi s)^{d/2}}{\rm e}^{-\frac{R^2}{4s}},
\end{equation} 
where $s$ is the proper time of the walk, i.e. $s=a^2\times{\,}({\rm
  number~ of~ steps})$. The probability obviously obeys the
conservation law $\int d^d{\cal R}P(s,R)=1$ and the initial condition
$\lim\limits_{s\to 0}^{}P(s,R)=\delta^{(d)}({\cal R})$. This 
probability is related to the proper-time representation of the
Green function $(-\partial^2)^{-1}_{{\cal R},0}=
\frac{\Gamma\left(\frac{d}{2}-1\right)}{4\pi^{d/2}R^{d-2}}
$ through 
\begin{equation}
\label{zerotemp}
 (-\partial^2)^{-1}_{{\cal R},0}=
\int_{0}^{\infty}\frac{ds}{(4\pi s)^{d/2}}{\rm e}^{-\frac{R^2}{4s}}.
\end{equation}

Let us generalize these formulae to temperatures below the
temperature of dimensional reduction. The probability then depends
also on the number $n$ of a Matsubara mode. In the same way as at
zero temperature, it can be obtained from the proper-time
representation of the Green function $(-\partial^2)^{-1}_{{\cal R},0}$. 
Instead of Eq.~(\ref{zerotemp}), one has (fixing from now on $d=4$):
\begin{equation}
\label{3}
(-\partial^2)^{-1}_{{\cal R},0}
\equiv\sum\limits_{n}^{}\int_0^\infty ds P_n=\sum\limits_{n}^{}\int_0^\infty\frac{ds}{(4\pi s)^2}
\exp\left[-\frac{{\bf R}^2+({\cal R}_4-\beta n)^2}{4s}\right],
\end{equation}
where $n$ in $\sum\limits_{n}^{}$ runs from $-\infty$ to $+\infty$,
and $\beta\equiv 1/T$. 
Note that this representation can be obtained directly
from~(\ref{1}) upon the decomposition $R^2\to{\bf R}^2+({\cal
  R}_4-\beta n)^2$. In the form~(\ref{3}), the
probability obviously obeys the conditions
$$\int d^4{\cal R}P_n(s,{\cal R})=1,~~ \lim\limits_{s\to 0}^{}P_n(s,{\cal R})=\delta^{(3)}({\bf R})
\delta({\cal R}_4-\beta n),$$ 
which are similar to those at zero temperature.

In case of a heavy quark-antiquark system, the random walk is not
free, as "the walker" is attached to the heavy quark $Q$ 
by the confining string. The length of the string is
$L=s/a$, i.e. the number of steps $s/a^2$ times the step size $a$.  Accordingly,
Eq.~(\ref{3}) becomes replaced by the Green function of the confined walker:
\begin{equation}
G({\bf R},T)\equiv\sum\limits_{n}^{}\int_{0}^{\infty} dsP_n(s,{\cal R})
{\rm e}^{-\beta\sigma s/a}=\sum\limits_{n}^{}\int_{0}^{\infty}\frac{ds}{(4\pi
s)^2}\exp\left[-\frac{{\bf R}^2+(\beta n)^2}{4s}- \frac{\beta\sigma s}{a}\right].
\end{equation} 
Here we use Eq.~(\ref{3}) with ${\cal R}_4=0$, since the quark and the antiquark at
the ends of the string are infinitely heavy. We will also use the zero-temperature value of the string tension 
$\sigma=(440{\,}{\rm MeV})^2$.

The full partition function of the random walk differs from this expression by the above-mentioned entropy factor, $N_c^{s/a^2}$, which should be included in the $s$-integration. The expression for the partition function is thus 
\begin{equation}
{\cal Z}({\bf R},T)=\sum\limits_{n}^{}\int_0^\infty\frac{ds}{(4\pi s)^2}\exp\left[-\frac{{\bf R}^2+(\beta n)^2}{4s}-
\frac{s}{a}\left(\sigma\beta-\frac{\ln N_c}{a}\right)\right].
\end{equation}
Performing the $s$-integration, we obtain
$${\cal Z}({\bf R},T)=\frac{1}{4\pi^2}\sqrt{\frac{1}{a}\left(\frac{\sigma}{T}-\frac{\ln N_c}{a}\right)}
\sum\limits_{n}^{}
\frac{K_1\Bigl(\sqrt{\frac{1}{a}\left(\frac{\sigma}{T}-\frac{\ln N_c}{a}\right)\left({\bf R}^2+(\beta n)^2
\right)}\Bigr)}{\sqrt{{\bf R}^2+(\beta n)^2}},$$
where $K_1$ is a Macdonald function. Next, to get the static $Q\bar Q$-potential, one should as usual take the 
limit of asymptotically large $Q\bar Q$-separations, $|{\bf R}|\to\infty$. There, the Macdonald function falls off 
exponentially, for which reason the sum can be approximated by its zeroth term alone. Furthermore, 
since the random walk is suppressed at $T<T_0$, its free energy vanishes at $T=T_0$, and therefore ${\cal Z}({\bf R},T)$
should be normalized by the condition ${\cal Z}({\bf R},T_0)=1$.
The effective string tension is then defined through the full free energy of the system, which is the sum of the usual linear potential and the normalized free energy of the random walk:
\begin{equation}
\label{sT}
\sigma(T)=\sigma-\frac{T}{R}\ln\frac{{\cal Z}(R,T)}{{\cal Z}(R,T_0)}\Biggr|_{R\to\infty}=\sigma+\frac{T}{\sqrt{a}}
\left(\sqrt{\frac{\sigma}{T}-\frac{\ln N_c}{a}}-\sqrt{\frac{\sigma}{T_0}-\frac{\ln N_c}{a}}\right).
\end{equation}
An estimate for $T_c$ now follows from the condition that the argument of the first square root vanishes:
\begin{equation}
\label{2n}
T_c\Bigr|_{N_c>1}=\frac{\sigma a}{\ln N_c}.
\end{equation}
Equating $T_c$ to the modern $N_c=3$ lattice value~\cite{latint1, lat}, 270 MeV, we obtain for the effective length of one string bit $a\simeq0.31{\,}{\rm fm}$. This is larger than the minimal possible value of this quantity, $a=0.22{\,}{\rm fm}$ -- the 
so-called vacuum correlation length~\cite{corrlength}, which defines the onset of a string-bit formation. 
As for the temperature $T_0$, it can be defined from the condition $\sigma(T_c)=0$, which yields
$$T_0=\frac{T_c}{\ln N_c+1}\simeq130{\,}{\rm MeV}.$$
An important finding of our model is the behavior 
\begin{equation}\label{find}
\sigma(T)\sim\sqrt{T_c-T}~~~ {\rm at}~~ T\to T_c.
\end{equation}
The same critical behavior follows from the 
Nambu-Goto model for the two-point correlation function of Polyakov loops~\cite{pa}. 

\begin{figure}
\epsfig{file=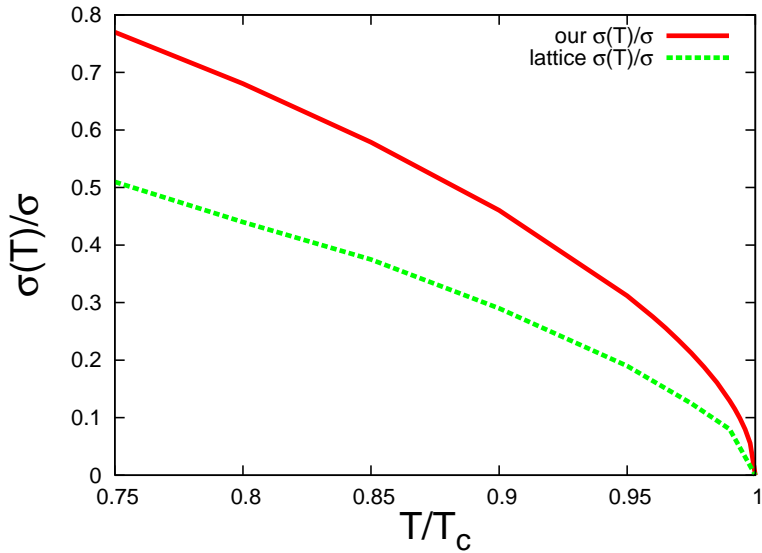, width=105mm}
\caption{\label{ra1} The ratio $\sigma(T)/\sigma$ at $N_c=2$ according to Eq.~(\ref{sT}) (full drawn curve) and Ref.~\cite{DFP} (dashed curve).}
\end{figure}

Let us consider the limiting case when string bits cannot alter color, i.e.
one should {\it formally} set in Eq.~(\ref{sT}) $N_c=1$, that yields
$$\sigma(T)=\sigma+\sqrt{\frac{\sigma T}{a}}\left(1-\sqrt{\frac{T}{T_0}}\right).$$
The fundamental difference of this limiting case from the realistic one $N_c=3$
is that here 
$\sigma(T)\sim(T_c-T)$ 
at $T\to T_c$. The corresponding value of the critical exponent $\nu=1$ defines the universality class of the {\it two-dimensional} Ising model, which
by no means can be realized in four-dimensional quenched SU($N_c$) QCD. 
The same linear fall-off of $\sigma(T)$ with $(T_c-T)$ one finds also 
in the Hagedorn phase transition and in the deconfinement scenario based on the condensation of long closed strings~\cite{revs, MO}. It comes as a mere consequence of the formula $\sigma(T)=\sigma-\frac{TS}{R}$ with the entropy $S\propto R$.
For long strings the entropy is just the logarithm of the number of possibilities to
realize on a lattice with the spacing $a$ a closed trajectory of length $L$.
Specifically for a hypercubic lattice, the entropy is 
$S=\frac{L}{a}\cdot\ln(2d-1)$. Therefore, the free energy of such a
closed string, $F=\sigma L-TS$, vanishes at $T_c=\frac{\sigma a}{\ln(2d-1)}$.  
Numerically, in $d=4$ dimensions, $T_c=103{\,}{\rm MeV}$, that
is by a large factor of 2.6 smaller than the modern lattice value 270~MeV. The principal
difference of our calculation is that we consider the random walk {\it between two different points}, taking into account the {\it confining force} between the final and the initial points of the walk along its trajectory. These are the two facts, which 
eventually lead to a different critical behavior. In conclusion of this section, we note that the idea to have a large entropy of the $Q\bar Q$-string in the vicinity of $T_c$ due to the valence gluons has recently been mentioned in Ref.~\cite{yuasim}.
However, the possibility that this mechanism leads to deconfinement has not been formulated in that paper.

\begin{figure}
\epsfig{file=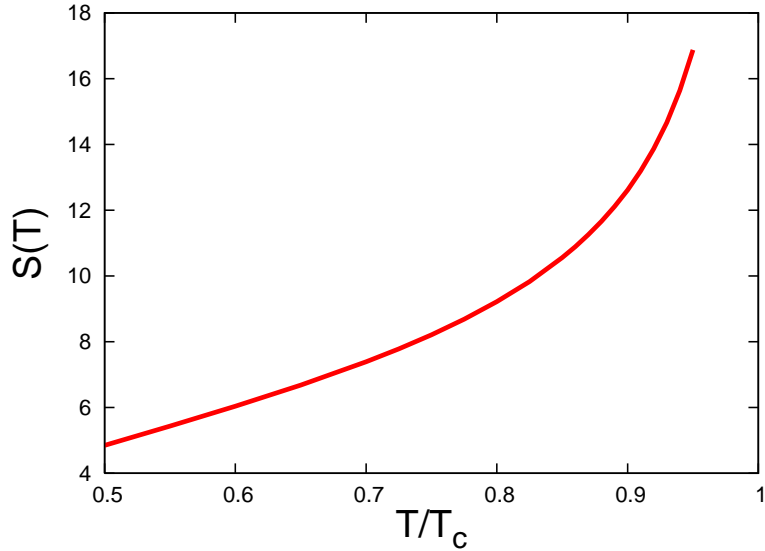, width=105mm}
\caption{\label{ra2} Entropy $S(T)$ of the static $Q\bar Q$-pair in the SU(2) quenched QCD at $R=1.5{\,}{\rm fm}$ and 
temperatures close to $T_c\simeq304{\,}{\rm MeV}$, 
calculated from Eq.~(\ref{sT}).}
\end{figure}

\begin{figure}
\epsfig{file=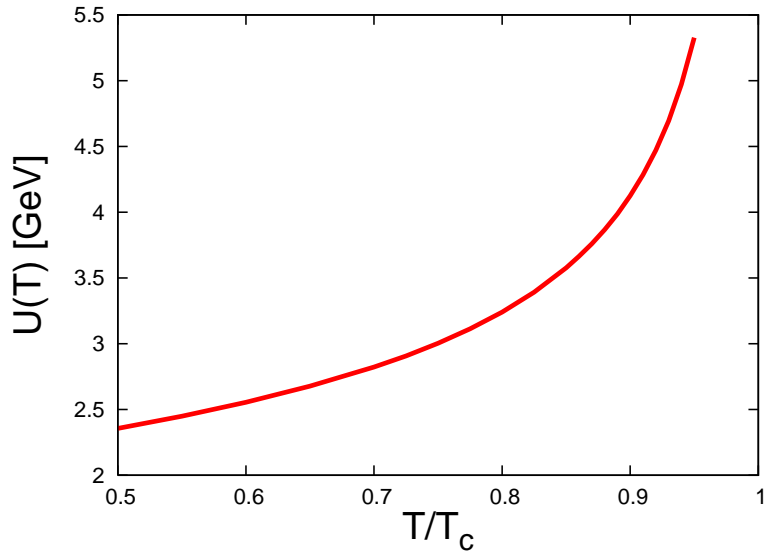, width=105mm}
\caption{\label{ra3} Internal energy $U(T)$ in GeV of the static $Q\bar Q$-pair in the SU(2) quenched QCD at $R=1.5{\,}{\rm fm}$ and temperatures close to $T_c\simeq304{\,}{\rm MeV}$, calculated from Eq.~(\ref{sT}).}
\end{figure}

It should finally be mentioned that, in the string models discussed in the previous paragraph, as well as in the model developed in this paper, the free energy vanishes {\it gradually} at $T\to T_c$. Therefore, none of these models
can describe first-order phase transition, which takes place in SU($N_c$) quenched QCD at $N_c\ge3$. They only  
can describe the phase transition in the SU(2) quenched QCD, which was studied on the lattice in Ref.~\cite{DFP}. In that case,
the mean-field critical exponent $\nu=1/2$, found in this paper and in Ref.~\cite{pa}, is changed to the 3d-Ising one, $\nu=0.63$.
This makes our values of $\sigma(T)$ larger than those obtained in Ref.~\cite{DFP}. Using, similarly to that paper, the value
$T_c=0.69\sqrt{\sigma}\simeq304{\,}{\rm MeV}$, we plot in Fig.~\ref{ra1} the ratio $\sigma(T)/\sigma$ following from Eq.~(\ref{sT}) at $N_c=2$. In the same figure, we present a curve interpolating the lattice data from Ref.~\cite{DFP}.
In Figs.~\ref{ra2} and~\ref{ra3}, we plot entropy and internal energy of the static $Q\bar Q$-pair in the SU(2) quenched QCD
at $R=1.5{\,}{\rm fm}$, calculated from Eq.~(\ref{sT}) by the formulae $S(T)=-R\frac{d\sigma(T)}{dT}$, $U(T)=\sigma(T)R+TS(T)$.
These quantities read 
$$S(T)=\sigma R\left[\frac{\beta}{2\sqrt{\sigma a(\beta-\beta_c)}}+\beta_c-\sqrt{\frac{\beta-\beta_c}{\sigma a}}\right]~~ 
{\rm and}~~ U(T)=\sigma R\left[1+\frac{1}{2\sqrt{\sigma a(\beta-\beta_c)}}\right],$$
respectively. In these equations, we use the value $\sigma a\simeq0.21{\,}{\rm GeV}$, which stems from Eq.~(\ref{2n}) upon the substitution $N_c=2$, $T_c=304{\,}{\rm MeV}$. This value of $\sigma a$ corresponds to $a\simeq0.21{\,}{\rm fm}$, which is again larger than the vacuum correlation length in the SU(2) quenched QCD, $0.16{\,}{\rm fm}$~\cite{CDM}.
Both $S(T)$ and $U(T)$ have a singularity
of the type $(T_c-T)^{-1/2}$ at $T\to T_c$. This singularity is weaker than $\delta\left(\frac{T_c-T}{T_c}\right)$, which takes 
place for the first-order phase transition at $N_c\ge 3$.

\section{Static $Q\bar Q$-pair with light quarks at $T<T_c$}
Let us now consider unquenched QCD with a heavy quark pair $Q\bar Q$
at large separation ($R\geq 1.5$ fm).  Then the $Q\bar
Q$-string breaks due to the production of a light $q\bar q$-pair. The
subsequent hadronization process leads to the formation of a
heavy-light meson ($\bar Qq$) or a heavy-light-light baryon ($Qqq$),
together with their antiparticles. We will calculate the thermodynamics of these
hadronic objects as a function of temperature. We consider the
($N_f=2$)-case, with light $u$- and $d$-quarks, and use the value
$T_c=200{\,}{\rm MeV}$ given in~\cite{tc} in order to compare with the
corresponding unquenched lattice simulations. Equation~(\ref{2n}) then
yields an effective value of $a=0.23{\,}{\rm fm}$, which is now used
in Eq.~(\ref{sT}) to determine $\sigma(T)$. For a heavy-light meson,
the Hamiltonian of the relativistic $q\bar Q$-system is
\begin{equation}
\label{H}
H_{\bar Qq}=m_{\bar Q}+\sqrt{{\bf p}^2+m_q^2}+V(r)~~ {\rm with}~~ V(r)=\sigma(T)r-(2-\delta)\sqrt{\sigma(T)}.
\end{equation}
The parameter $m_{\bar Q}$ denotes the heavy-antiquark mass, $m_q$ is the constituent mass of the light quark, $m_q\simeq 300{\,}{\rm MeV}$. 
The subtraction of $2\sqrt{\sigma(T)}$ in $V(r)$ is known to be important to reach agreement between the predictions of the relativistic quark model with the phenomenology of meson spectroscopy~\cite{const2, const1}. 
An additional correction $\delta\sqrt{\sigma(T)}$ ensures that the mass of the lowest state in our calculation corresponds to the 
lightest heavy-quark meson. We use the $D^0$-meson with
$m_{D^0}=1.864{\,}{\rm GeV}$ and $m_{\bar Q}=m_c=1.48{\,}{\rm GeV}$ to fix the parameters of our model.

The square root in the Boltzmann factor
can be handled by  integrating over an auxiliary parameter:
$$\exp\left(-\beta \sqrt{{\bf p}^2+m_q^2}\right)=\frac{2}{\sqrt{\pi}}
 \int_0^\infty d\mu \exp\left(-\mu^2-\frac{\beta^2}{4\mu^2}\left({\bf p}^2+m_q^2\right)\right).$$
One then arrives at a 3d Schr\"odinger equation of the form
$$
(-K\vec\nabla^2+M|{\bf x}|)\psi({\bf x})=E\psi({\bf x}),$$
where $K$ and $M$ are positive constants of dimension $[{\rm length}]$ and $[{\rm mass}]^2$, respectively. 
Its eigenenergies read~\cite{const2}
$$E_{n_r,l}= \alpha_{n_r,l}\, (K M^2)^{1/3},$$
where $\alpha_{n_r,l}$'s are positive numbers with $\alpha_{00}\simeq 2.34$. In calculating  the 
partition function of the heavy-light mesons $(\bar Qq$ and $Q\bar q)$~\cite{const2} we take their ground state into account exactly and model their higher 
eigenenergies by those of a 3d harmonic oscillator with the frequency 
$$\omega\equiv(K M^2)^{1/3}=\left(\frac{\beta\sigma^2(T)}{4\mu^2}\right)^{1/3}.$$
This strategy works well for the lowest states, and higher states are Boltzmann-suppressed, so the lack of accuracy 
does not matter so much. Because of the 
degeneracy factor $(n/2+1)(n+1)$ for the oscillator, we have 
$$
{\cal Z}_{{\rm 2 mes}}=4{\cal Z}_{\bar Qq}^2~~$$ 
with 
$${\cal Z}_{\bar Qq}={2\over \sqrt{\pi}}\exp\left(-\beta m_{\bar Q} + (2-\delta)\beta \sqrt{\sigma(T)}\right)\times$$
\begin{equation}
\label{11}
\times\sum_{n=0}^\infty
\left(\frac{n}{2}+1\right)(n+1)\int_0^\infty d\mu
\exp\left[-(n+\alpha_{00})\beta\omega-\mu^2-\frac{\beta^2 m_q^2}{4\mu^2}\right].
\end{equation}
The partition function of two noninteracting mesons is $4{\cal
Z}_{\bar Qq}^2$, where the factor 4 is a product of two flavors and
two spin states of the mesons because the spins and isospins of the
produced $q$ and $\bar q$ are coupled to isospin 0 and spin 0 by string breaking. 
The mass $m_{\bar Q}$ of the heavy antiquark is dropped
in Eq.~(\ref{11}) in order to compare our calculation with the lattice data \cite{tc},
which simulate a heavy quark and antiquark by two Polyakov lines.  Doing the sum
over $n$ analytically, we arrive at the following result:
\begin{equation}
\label{fe}
F_{{\rm 2 mes}}=-T \ln\frac{16}{\pi}- (4-2\delta)\sqrt{\sigma(T)}
-2 T \ln\Biggl[\int_0^\infty d\mu
\exp\left(-\mu^2 - 
\frac{\beta^2 m_q^2}{4 \mu^2}\right) \frac{\exp(-\alpha_{00}\beta\omega)}{(1-\exp(-\beta\omega))^3}\Biggr].
\end{equation}
The remaining $\mu$-integration is done numerically. The
necessary value of the parameter $\delta$ to reproduce the correct $m_{D^0}$
mass is $\delta=0.354$, which can be obtained from the free energy $F$ in the
limit $T\to0$. Entropy and internal energy can be calculated by the
standard formulae $S=-\partial F/\partial T$, $U=F+TS$. Note that,
when calculating the entropy, the derivative $\partial/\partial T$
should not act on the Hamiltonian in the partition function, since
otherwise the consistency of the thermodynamic relations would be
violated.

It is also possible that string breaking generates a baryon and an
antibaryon instead of two mesons. The Hamiltonian for a baryon reads
$$H_{Qqq}=m_Q+\sum\limits_{i=1}^{2}\left(\sqrt{{\bf
p}_i^2+m_q^2}+V(r_i)\right),$$ where we approximate the position of
the baryon string junction by the position of the heavy quark $Q$,
which is legitimate due to the heaviness of $Q$. Since the heavy quark
$Q$ is static, terms from the two light quarks separate and one can
deal with this Hamiltonian similar to the meson case. The
corresponding free energy of baryon and antibaryon reads
\begin{equation}
F_{{\rm 2 bar}}=-T \ln\frac{16}{\pi^2}- (8-4\delta)\sqrt{\sigma(T)}
-4 T \ln\Biggl[\int_0^\infty d\mu
\exp\left(-\mu^2 - 
\frac{\beta^2 m_q^2}{4 \mu^2}\right) \frac{\exp(-\alpha_{00}\beta\omega)}{(1-\exp(-\beta\omega))^3}\Biggr].
\end{equation}
 For simplicity, we restrict ourselves to diquark quantum numbers with
 zero isospin and zero spin.  The parameter $\delta$ in the baryon
 calculation is adjusted so that the corresponding ground state
 corresponds to the lightest charmed baryon $\Lambda_c^+$ with
 $m(\Lambda_c^+)=2.286{\,}{\rm GeV}$. The resulting  value $\delta=0.393$ is not
 very different from $\delta=0.354$ obtained for the heavy-meson system, therefore we use the averaged
 value $\delta=0.37$ in both cases.

\begin{figure}
\epsfig{file=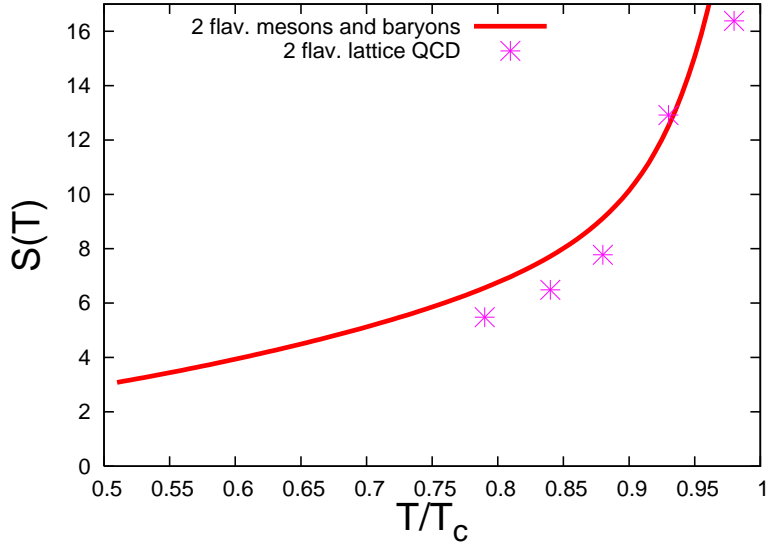, width=105mm}
\caption{\label{Su} The calculated entropy $S(T)$  (full drawn curve) of two
  mesons and two baryons produced by string breaking for a heavy $Q$ and
  $\bar Q$ at large separation ($R\ge1.5$ fm) is shown as a function of $T/T_c$ with $T_c=200$ MeV. 
The stars show the lattice
  data~\cite{tc} for the same $Q\bar Q$ configuration. }
\end{figure}

\begin{figure}
\epsfig{file=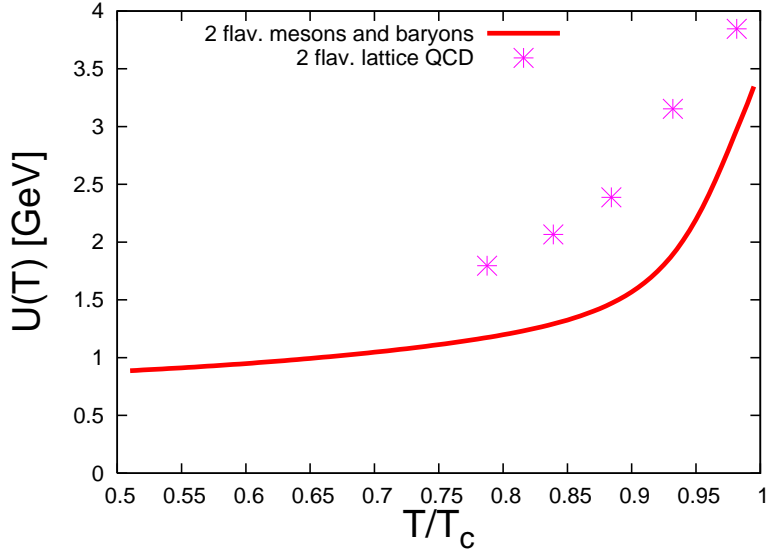, width=105mm}
\caption{\label{Uu} The calculated internal energy $U(T)$ (full drawn curve) of two
  mesons and two baryons produced by string breaking for a heavy $Q$ and
  $\bar Q$ at large separation ($R\ge1.5 $ fm) is shown as a function of $T/T_c$ with $T_c=200$ MeV. 
The stars show the lattice
  data~\cite{tc} for the same $Q\bar Q$ configuration.}
\end{figure}

Combining the contributions to the partition function coming from mesons and baryons, we have
\begin{equation}\label{Ftot}
F=-T \ln(Z_{\rm 2 mes}+Z_{\rm 2 bar})=-T\ln\left(4 {\cal Z}_{\bar Qq}^2+ {\cal Z}_{\bar Qq}^4  \right).
\end{equation}
In Figs. \ref{Su} and \ref{Uu}, we plot the entropy and the internal
energy corresponding to Eq.~(\ref{Ftot})  together with the
corresponding lattice data ~\cite{tc}. Here $S(T)$ and $U(T)$ mean the
change in entropy and internal energy due to the presence of a heavy
quark-antiquark pair in the hadronic heat bath. The heat bath 
justifies the calculation in the canonical ensemble in our case. The heavy
$Q$ and $\bar Q$ and the two light $q$ and $\bar q$ from string breaking  are the sources of our
mesons. Therefore we think that a comparison of our model calculation with
the lattice simulation is appropriate.  The agreement is very
good for the entropy, while the internal energy coming from our
calculations is smaller than the lattice results. Our curve is shifted 
downwards by $\Delta U\simeq-0.5{\,}{\rm GeV}$ compared with the lattice simulations. 
Due to the simplicity of our calculation the qualitative agreement is
nevertheless good.

\section{Static $Q\bar Q$-pair with light quarks at $T>T_c$}

In this section we consider a heavy $Q\bar Q$ pair at temperatures
$T\ge T_c$, i.e. after deconfinement with free quarks and gluons present in
the plasma.  The heavy quark and antiquark are immersed into this
plasma at the points ${\bf x}_Q$ and ${\bf x}_{\bar Q}$ at large
separation ($R\geq 1.5$ fm).  The non-Abelian interaction energy of a
plasma constituent $a$ with the $Q\bar Q$ pair has the following form:
\begin{equation}\label{potenergie}
{\cal V}^a({\bf r})=    C^{aQ} {\cal U}(\vert{\bf r}- {\bf x}_{Q}\vert) + C^{a\bar Q}{\cal U}(\vert{\bf r}- {\bf x}_{\bar Q}\vert).
\end{equation}
Here, $C^{aQ}$ and $C^{a\bar Q}$ are the corresponding color factors, namely products of SU(3) generators, which are discussed in Appendix~\ref{cf}, and the interaction potential ${\cal U}({\bf r})$ is given by a screened gluon exchange between the colored sources
\begin{equation}\label{Uww}
{\cal U}({\bf r})=4\pi\int \frac{d^3{\bf k}}{(2\pi)^3} \frac{\alpha_s(k,T)}{k^2+m_D^2}{\rm e}^{i {\bf k}{\bf r}}.
\end{equation}

For $\alpha_s$, we use a running coupling constant at finite temperature $T$ which has been derived in Ref.~\cite{bpir, bgies}
\begin{eqnarray}
\alpha_s(k,T)&=& \frac{u_1 \frac{k}{T}}{1+\exp\left(u_2\frac{k}{T}-u_3\right)}+\nonumber\\ &+& \frac{v_1}{\left(1+\exp(v_2\frac{T}{k} - v_3)\right) \ln\left({\rm e} + \left(\frac{k}{\Lambda_s}\right)^a + \left(\frac{k}{\Lambda_s}\right)^b \right)}.
\end{eqnarray}
This running coupling increases for small momenta $k$, has a maximum at $k \simeq T$ and finally decreases for $k\to \infty$. 

The Debye mass is
\begin{equation}
\label{MD}
 m_D=\sqrt{\frac{N_c}{3}+\frac{N_f}{6}}\,\, gT~~ \mbox{with}~~ g=\sqrt{4\pi\alpha_s(m_D,T)}\simeq2.5.
\end{equation}
This $g$ is determined from the selfconsistency equation \cite{bpir} with the running coupling. 

The interaction energy~(\ref{potenergie}) is to be summed over all the $N$ constituents $a$ of the plasma, ${\cal V}({\bf r}_1,\ldots,{\bf r}_N)=\sum_{i=1}^N {\cal V}^{a_i}({\bf r}_i)$.
The perturbative expansion of the free energy in classical thermodynamics reads (see~\cite{landau})
\begin{eqnarray}\label{F}
F=F_0+F_{Q\bar Q}+\langle {\cal V}({\bf r}_1,\ldots,{\bf r}_N)\rangle -{1\over 2T}\left(\langle {\cal V}^2({\bf r}_1,\ldots,{\bf r}_N)\rangle -\langle {\cal V}({\bf r}_1,\ldots,{\bf r}_N)\rangle ^2\right),
\end{eqnarray}
where for each operator 
$$\left<{\cal O}({\bf r}_1,\ldots,{\bf r}_N)\right>={\rm Tr}_1\left(\prod\limits_{i=1}^{N}
\int\frac{d^3{\bf p}_id^3{\bf r}_i}{(2\pi)^3}\right){\cal O}({\bf r}_1,\ldots,{\bf r}_N){\rm e}^{\beta(F_0-E_0)}.$$
Here, $F_{Q\bar Q}$ is the free energy of the interacting heavy $Q$ and $\bar Q$, 
$F_0$ is the free energy of the quark-gluon plasma, and the rest describes the interaction of the plasma constituents 
with the $Q\bar Q$-pair.  
Tr$_1$ denotes the trace over color indices of the interacting plasma constituent in the $Q\bar Q$ singlet state. The first-order term $\langle V({\bf r}_1, \dots, {\bf r}_N)\rangle$ vanishes due to color neutrality of the plasma, the second-order term, however, yields a nonvanishing result and describes the various interactions of the plasma constituents with the $Q\bar Q$ singlet state. The respective color factors can be calculated as products of generators in the various representations of SU(3). In second order, only the diagrams with {\it two} interactions of {\it one} plasma constituent with either $Q$ or $\bar Q$ or both of them are nonzero.

The energy of $N_q$ light quarks, $N_{\bar q}$ light antiquarks
and $N_g$ gluons is $E_0=\sum^N_{i=1} \sqrt{{\bf p}_i^2+m_i^2}$, where $N=N_q+N_{\bar q}+N_g$. Here the squared mass of quarks and antiquarks is~\cite{peshier, lebellac}
\begin{equation}
m_q^2=m_{\bar q}^2=m_0^2+ 2 gT \sqrt{\frac{N_c^2-1}{16 N_c}}\left(m_0+gT \sqrt{\frac{N_c^2-1}{16 N_c}}\right)
\end{equation}
and the kinetic mass of gluons is given by  $m_g^2= \frac12 m_D^2 $ \cite{peshier, lebellac} at $N_c=3$.
We also use $g=2.5$ as above and set the current quark mass to $m_0=30$~MeV.

Although our heuristic introduction used classical statistics we will
calculate in the following the free energy in quantum statistics as it is appropriate. 
The zeroth--order
contribution $F_0$ reads
\begin{eqnarray}\label{F0}
F_0&=&-\sum\limits_{a=q,\bar q,g}^{} N_a T \ln\left[ \int  d^3{\bf p}\, d^3{\bf r} \exp\left(-\beta \sqrt{{\bf p}^2 +m_a^2}\right)                              \right]\nonumber\\
&=&-\sum_a N_a T \ln\left(4 \pi V T m_a^2 K_2\left({m_a\over T}\right)\right),
\end{eqnarray}   
where $K_2$ is a Macdonald function. The contribution $F_0$ of the quark-gluon plasma alone 
without the heavy   $Q\bar Q$ pair is subtracted in lattice calculations~\cite{lat, tc}.
Therefore we will also leave it out in our results. 
 
The contribution of the $Q\bar Q$-interaction,
$$F_{Q\bar Q}(R, T)=-\frac{4}{3}\alpha_s\frac{{\rm e}^{-m_D R}}{R},$$ 
can be neglected because it is of the order of a few MeV at large distances $R\simeq 1.5{\,}{\rm fm}$.

The term  $\langle {\cal V}^2({\bf r}_1,\ldots,{\bf r}_N)\rangle$ in Eq.~(\ref{F}) is associated with two interactions of each plasma constituent. The contribution is the same for constituents of the same kind: 
\begin{eqnarray}\label{F2}
F_2&=&-{1\over 2T} \langle {\cal V}^2({\bf r}_1,\ldots,{\bf r}_N)\rangle\nonumber\\ 
&=& - \frac{2}{\pi T} \left(\frac29 n_q^{{\rm eff}}+\frac29 n_{\bar q}^{{\rm eff}}+ \frac12 n_g^{{\rm eff}}\right) \int d^3 {\bf q} \frac{\alpha_s(q, T)^2 }{(q^2+m_D^2)^2}\left(1- {\rm e}^{i {\bf q}( {\bf x}_Q- {\bf x}_{\bar Q})}   \right)
\end{eqnarray}
(see Appendix \ref{app:F2}).  In this equation, the color factors $c_q=c_{\bar q}=2/9$
and $c_g=1/2$ appear multiplying the effective densities $n_a^{{\rm eff}}=
N_a^{{\rm eff}}/V$ of quarks, antiquarks or
gluons, respectively. Since the heavy  $Q\bar Q$ pair is color neutral, the contribution $F_2$
vanishes, when the  $Q\bar Q$ - distance $|{\bf x}_Q- {\bf x}_{\bar Q}|$ approaches zero.
The running of the coupling with temperature is
crucial for the strong decrease of the entropy and internal energy
with temperature. For a fixed value of $\alpha_s$ we would have
\begin{equation}
\label{FRen}
 F_2=-\frac{1}{2T} \left(\frac29 n_q^{{\rm eff}}+
\frac29 n_{\bar q}^{{\rm eff}}+ \frac12 n_g^{{\rm eff}}\right) \alpha_s^2 {4\pi\over m_D}\left(1-{\rm e}^{-m_D R}\right),~~ R=|{\bf x}_Q-{\bf x}_{\bar Q}|=1.5{\,}{\rm fm},
\end{equation}
and the corresponding entropy and internal energy would decrease with temperature even slower.

The next step is to calculate the effective densities, starting from the free ones.
The particle densities of free quarks and gluons are listed in Appendix~\ref{ns}, and the results are
\begin{eqnarray}
\label{densities}
&n_q&=n_{\bar q}=3 T N_f \frac{m_q^2}{\pi^2}  \sum_{n=1}^\infty \frac{(-1)^{n+1}}{n} K_2\left({m_q\over T} n \right),\nonumber\\
&n_g&=12 T \frac{m_g^2}{\pi^2}  \sum_{n=1}^\infty \frac{1}{n} K_2\left({m_g\over T} n \right) . 
\end{eqnarray}
The full density is a sum of the three contributions $n_q+n_{\bar q}+n_g$. Confinement requires 
$n_{g,q, \bar q}^{\rm eff}(T_c)=0$, therefore we add a factor $h(T)$ in the effective densities:  
\begin{equation}
\label{fulldens}
n^{{\rm eff}}_a(T)=h(T)n_a
\end{equation}
with
\begin{equation}\label{f}
h(T)=1-\exp\left(- \frac{T-T_c}{\lambda}\right).
\end{equation}

\begin{figure}
\epsfig{file=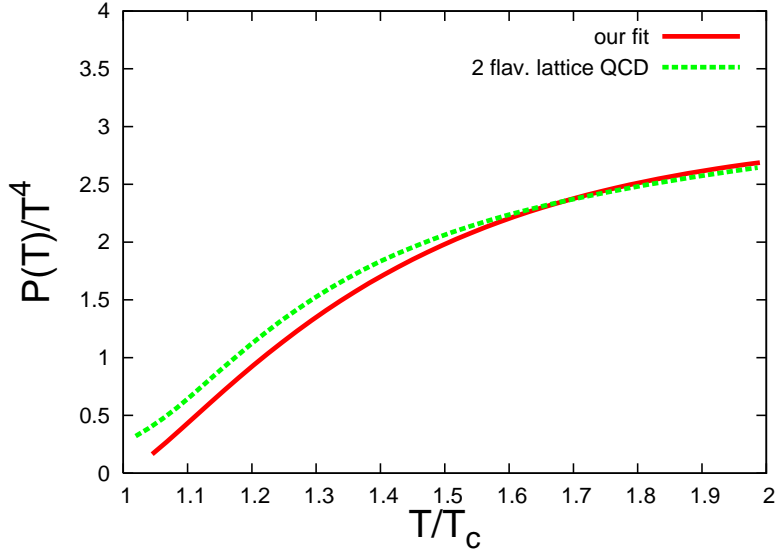, width=105mm}
\caption{\label{poverpinf}
The calculated ratio $P_{{\rm eff}}(T)/T^4$
of Eq.~(\ref{Peff}) (full drawn curve) is shown in comparison with
  the one from lattice simulations of Ref. \cite{karsch} (dashed curve) 
for two light flavors. }
\end{figure}

To obtain the parameter $\lambda=60$ MeV  in $h(T)$, we have calculated the
pressure corresponding to these effective densities and
compared it with the pressure from lattice simulation in Ref.~\cite{karsch}.  
The grand canonical partition function from quarks, antiquarks
and gluons has the form:
\begin{eqnarray}
\ln Z_{\rm grand}= \sum_a D_a \frac{V T m_a^2}{2 \pi^2}K_2\left({m_a\over T}\right).
\end{eqnarray} 
Here, $D_a$ are the corresponding degeneracy factors, $D_q=D_{\bar
  q}=N_c N_f (2s_q+1)$ and $D_g=(N_c^2-1) N_p$ which contain $N_p=3$ as the number
of polarizations of a massive gluon, and $s_q=1/2$ as the spin of the
quark.

The pressure of free particles
\begin{equation}\label{pressure}
P={T\over V} \ln Z_{\rm grand}=\frac{6T^2}{\pi^2}\left(2m_g^2 K_2\left(\frac{m_g}{T}\right)
+ \frac{N_f}{2} m_q^2K_2\left({m_q\over T}\right)+ \frac{N_f}{2} m_{\bar q}^2K_2\left({m_{\bar q}\over T}\right)\right) 
\end{equation}
would not vanish below $T_c$. To have a vanishing pressure at $T=T_c$,
we also modify it by the confinement factor $h(T)$. Massless quarks
and gluons have the density $n\propto T^3$ and the pressure $P\propto
T^4$. Therefore, $P_{{\rm eff}}(T)$ is approximately suppressed by the
factor $h(T)^{4/3}$:
\begin{equation}\label{Peff}
P_{{\rm eff}}(T)=h(T)^{4/3}\cdot\frac{6T^2}{\pi^2}\left(2 m_g^2 K_2\left({m_g\over T}\right)+\frac{N_f}{2} m_q^2 K_2\left({m_q \over T}\right)+\frac{N_f}{2} m_{\bar q}^2 K_2\left({m_{\bar q} \over T}\right)\right).
\end{equation}
In Fig.~\ref{poverpinf}, we plot the ratio $P(T)/T^4$ from Ref.~\cite{karsch}
and the ratio $P_{\rm eff}(T)/T^4$ from Eq.~(\ref{Peff}). 

\begin{figure}
\epsfig{file=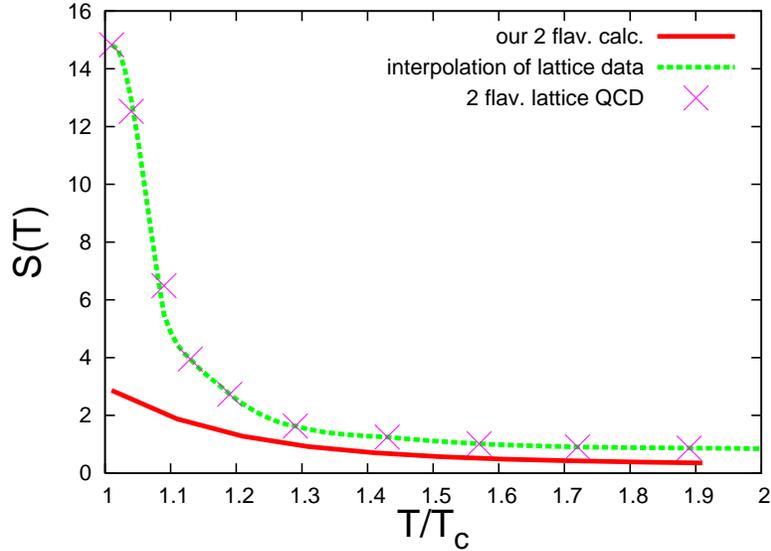, width=105mm}
\caption{\label{S} The calculated net entropy $S(T)$ at $N_f=2$ (full drawn curve) 
of a $Q\bar Q$ pair at large separation is shown as a function of $T/T_c$ with $T_c=200$ MeV. 
The entropy of the plasma without
the  $Q$ and $\bar Q$ pair is subtracted. Our calculation
is compared with a function interpolating  the lattice data \cite{tc, lat} 
(dashed curve) for better visibility.}
\end{figure}

\begin{figure}[t]
 \epsfig{file=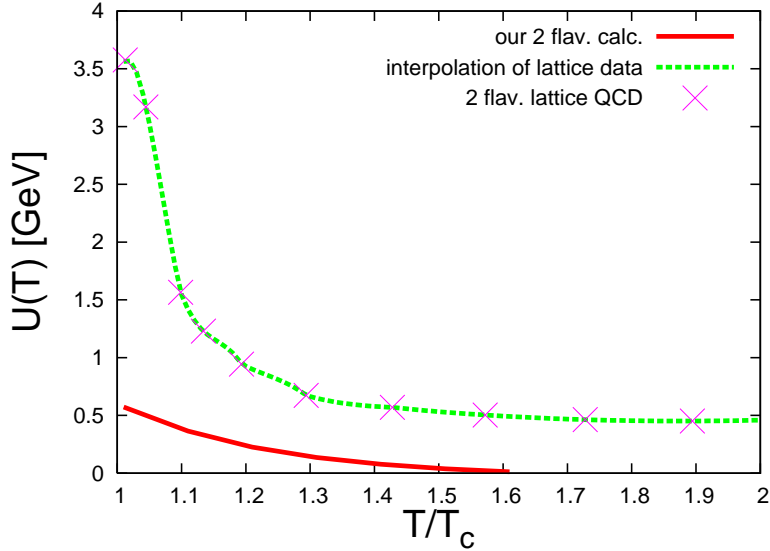, width=105mm}
 \caption{\label{U} The calculated  net internal energy $U(T)$ in GeV (full drawn curve) 
of a $Q\bar Q$ pair at large separation  is shown as a function of $T/T_c$ with $T_c=200$ MeV.
The internal energy of the plasma without
the  $Q$ and $\bar Q$ pair is subtracted. Our calculation 
is compared to a function interpolating the lattice data \cite{tc, lat} 
(dashed curve) for better visibility.}
\end{figure}

In the second-order free energy $F_2$, Eq.~(\ref{FRen}), we include the effective densities $n_a^{{\rm eff}}$,  
Eq.~(\ref{fulldens}), modified by $h(T)$.
In Fig. \ref{S} and Fig. \ref{U}, we have plotted the entropy $S$ and the internal energy $U$ following from the standard thermodynamic relations $S(T)=-\partial F_2 / \partial T$ and $U(T)=F_2(T)+T S(T)$ for two flavors. For the critical temperature we choose $T_c=$200 MeV for two flavors~\cite{tc}.

\begin{figure}
\epsfig{file=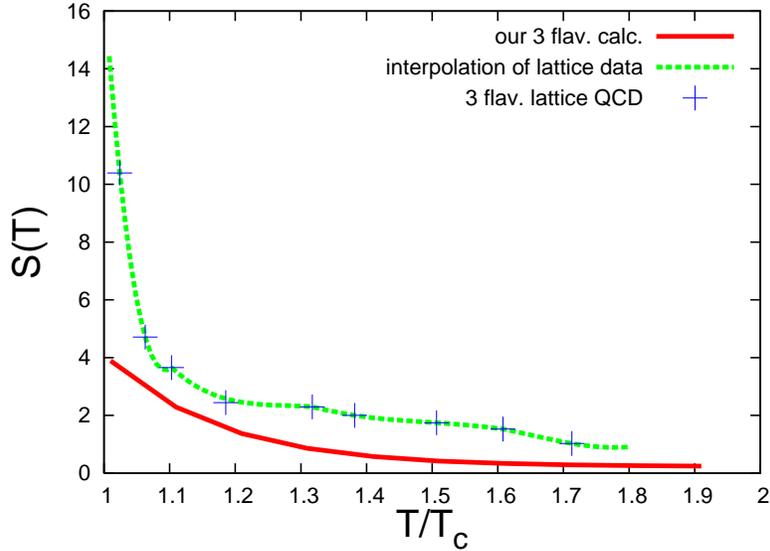, width=105mm}
\caption{\label{nf1} The calculated net entropy $S(T)$ at $N_f=3$ (full drawn curve) 
of a $Q\bar Q$ pair at large separation is shown as a function of $T/T_c$ with $T_c=193$ MeV. 
The entropy of the plasma without
the  $Q$ and $\bar Q$ pair is subtracted. Our calculation
is compared with a function interpolating  the lattice data \cite{tc, lat} 
(dashed curve) for better visibility.}
\end{figure}

\begin{figure}[t]
 \epsfig{file=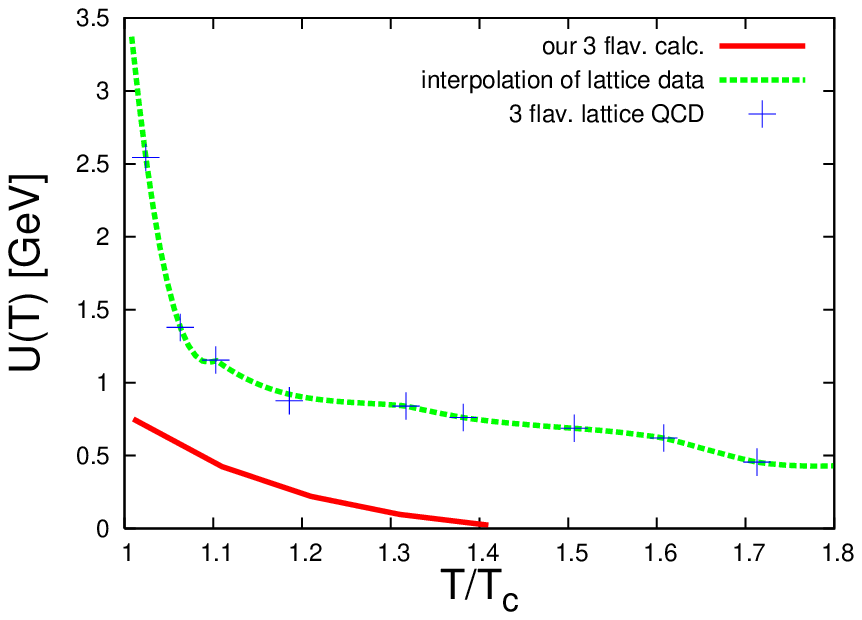, width=105mm}
 \caption{\label{nf2} The calculated  net internal energy $U(T)$ in GeV at $N_f=3$ (full drawn curve) 
of a $Q\bar Q$ pair at large separation  is shown as a function of $T/T_c$ with $T_c=193$ MeV.
The internal energy of the plasma without
the  $Q$ and $\bar Q$ pair is subtracted. Our calculation 
is compared to a function interpolating the lattice data \cite{tc, lat} 
(dashed curve) for better visibility.}
\end{figure}

As one can see, $S(T)$ and $U(T)$ above $T_c$ are much smaller than in  
Refs.~\cite{tc, lat}. We observe a difference $\Delta U\simeq-0.5{\,}{\rm GeV}$ for temperatures larger than $1.1 T_c$ between
our values of the internal energy and those of the lattice data,
similarly to the result of the unquenched calculation at $T<T_c$. 
The trend from the lattice simulations that $S(T)$ and $U(T)$ decrease with $T$, is also seen here.
For the case of three quark flavors, we show the results in Figs.~\ref{nf1}, \ref{nf2}.
The calculation of the entropy for three flavors is close to
the one for two flavors. A similar result is obtained in the lattice simulations.
However, there is one qualitative difference: Near $T_c$ two flavor lattice simulations give a larger $S(T)$ and $U(T)$ than simulations with three flavors. Our calculation yields
opposite results. They arise from the flavor dependence of the
densities and the Debye mass, which partially cancel and lead to an 
approximate independence of $S(T)$ and $U(T)$ on the number of flavors. 

We can enhance the internal energy and entropy by modifying our system 
in such a way that all quarks and antiquarks are bound in color-octet states and interact as such with the 
$Q\bar Q$-pair~\cite{shu:bound}. This yields an estimate for the influence of bound states on our calculations.
 
\begin{figure}
\begin{center}
\epsfig{file=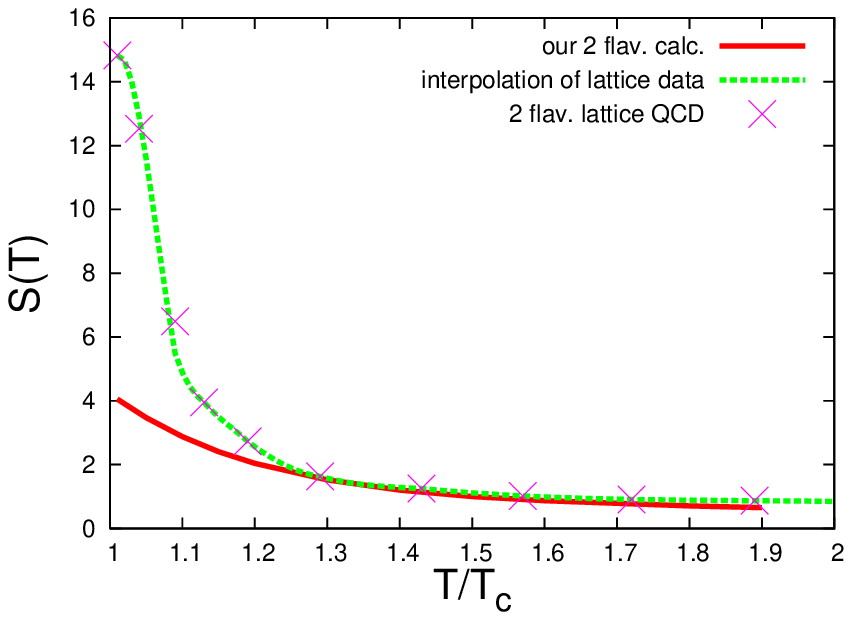, width=105mm}
\caption{\label{S8} The calculated net entropy $S(T)$ (full drawn curve)
  of a $Q\bar Q$ pair at large separation interacting with $q\bar
  q$ bound states and gluons is shown as
  a function of $T/T_c$ with $T_c=200$ MeV.  The entropy of the plasma
  without the $Q$ and $\bar Q$ pair is subtracted. Our calculation
   is compared to a function interpolating the lattice
  data \cite{tc, lat} (dashed curve) for better
  visibility.}
\end{center}
\end{figure}

\begin{figure}
\begin{center}
\epsfig{file=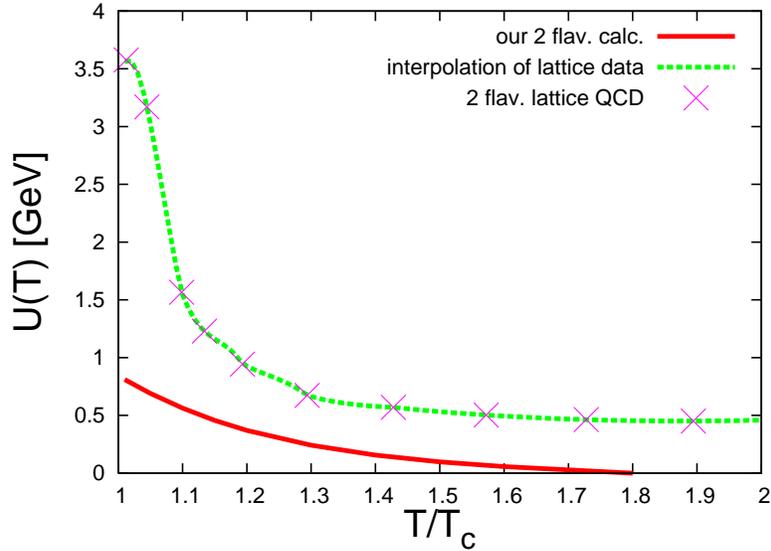, width=105mm}
\caption{\label{U8}  The calculated net internal energy $U(T)$ in GeV (full drawn curve)
  of a $Q\bar Q$ pair at large separation interacting with $q\bar
  q$ bound states and gluons is shown as
  a function of $T/T_c$ with $T_c=200$ MeV.  The internal energy of the plasma
  without the $Q$ and $\bar Q$ pair is subtracted. Our calculation 
   is compared to a function interpolating the lattice
  data \cite{tc, lat} (dashed curve) for better
  visibility.}
\end{center}
\end{figure}

For simplicity we assume that the $q\bar q$ bound states have a mass of 2$m_q$. The corresponding degeneracy factor is now $(N_c^2-1)N_f^2 N_s=16 N_f^2$ ($N_s=2$ for spin 0 or 1). Therefore, the density is given as (cf. Eq. (\ref{densities}))
\begin{equation}
n_{q\bar q}= \frac{32}{\pi^2}T N_f^2 m_q^2 \sum_{n=1}^\infty \frac1n K_2\left(\frac{2m_q}{T} n\right).
\end{equation}
As before, we obtain the effective density through the multiplication by $\widehat{h}(T)$, where the constant $\widehat{\lambda}$ is again fitted to reproduce the pressure, similarly to Eq. (\ref{Peff}). This yields $\widehat{\lambda}=80$ MeV for two flavors. The color factor $c_{q\bar q}=1/2$ is the same as for gluons, and one can again determine entropy and internal energy from Eq. (\ref{F2}). These quantities are shown in Figs. \ref{S8} and \ref{U8}. 

In the vicinity of $T_c$ the entropy remains smaller than in the lattice calculations, but for higher temperatures it is in good agreement with the lattice data. In contrast, the internal energy is not very much different from the one without $q\bar q$ bound states. This shows that our calculation of the internal energy in second order perturbation theory is not changed very much by the presence of bound states in contrast to the otherwise~\cite{shu:bound, k:bound, c:bound} 
important effects of bound states in the quark-gluon plasma.

\section{Summary}
In the present paper, we have analytically addressed the
thermodynamics of a static quark-antiquark pair at large separation
$R\geq 1.5$ fm in the vicinity of the deconfinement phase transition.
In quenched QCD for $T<T_c$, the quark-antiquark string passes through
valence gluons and forms a gluon chain. We have derived the resulting
temperature-dependent string tension, Eq.~(\ref{sT}), which vanishes
at the critical point as $\sqrt{T_c-T}$. In unquenched QCD, for $
T<T_c$ heavy-light mesons and heavy-light-light baryons are produced
due to string breaking. The internal energy and entropy of our
calculation fit quite well the corresponding lattice data. Finally, we
have calculated the interaction energy for $T>T_c$ in unquenched QCD
of the Debye-screened static quark-antiquark pair with the constituents of
the plasma in second order thermodynamic perturbation theory. The
change of the free energy depends on the effective density of 
quarks and gluons in the plasma, which has been constrained to vanish
at $T=T_c$ and is adjusted in such a way that the pressure fits the one
in lattice QCD for $N_f=2$.  The entropy and internal energy of the model calculation differ from 
the lattice data. The large entropy near $T_c$ cannot be reproduced. 
The internal energies $U(T)$
calculated for the heavy $Q\bar Q$ pair are in both cases $T<T_c$ and
$T>T_c$ by about 0.5--0.6 GeV lower than in the lattice simulations. The trend of a falling $S(T)$ and $U(T)$ with increasing $T$ 
is also seen in our model calculation. The heavy $Q\bar Q$-pair, immersed at large separation in the 
quark-gluon plasma, becomes less and less relevant at higher temperatures.
Finally, the influence of possible $q\bar q$ bound states on our
calculations has been estimated. Qualitatively, they do not change our
results very much. If one wants to answer the question whether the 
hot QCD system  is strongly interacting, our answer is twofold:
Below $T_c$ the gluonic string and the excited hadronic states definitely 
represent strongly interacting composite systems. Above $T_c$ we have the screened
color Coulomb-potential with its strength enhanced by the running QCD-coupling.
The resulting calulated entropy, however, cannot fully reproduce  the entropy
from lattice simulations. At large temperatures the agreement becomes better.
Surprisingly the internal energy of the simplified calculation and the lattice
simulations do not converge. This divergence should be investigated more thoroughly. 

\acknowledgments
\noindent
The work of D.A. has been supported through the contract MEIF-CT-2005-024196. S.D. thanks F.~Karsch, P.~Petreczky and F.~Zantow 
for providing him with the details of the lattice data.

\begin{appendix}

\section{Computation of the sum for the three-dimensional harmonic oscillator}
The sum over $n$ is computed by means of the formula
\begin{equation}
\sum_{n=0}^\infty n \exp(-\lambda \beta\omega n)= -{1\over \beta\omega} \frac{\mbox{d}}{\mbox{d}\lambda} \sum_{n=0}^\infty \exp(-\lambda \beta\omega n)=
 -{1\over \beta\omega} \frac{\mbox{d}}{\mbox{d}\lambda} \frac{1}{1-\exp(-\lambda \beta\omega)}
\end{equation}
for the geometric series. Therefore, every factor of $n$ in the sum can be replaced by $-{1\over \beta\omega}
\frac{d}{d\lambda}$, and we have
\begin{eqnarray}\label{sum}
\sum_{n=0}^\infty \left(\frac{n}{2}+1\right)(n+1){\rm e}^{-\lambda \beta\omega n} 
 &=&\left(\frac{1}{2(\beta\omega)^2}
\frac{\mbox{d$^2$}}{\mbox{d}\lambda^2} -\frac{3}{2\beta\omega}  \frac{\mbox{d}}{\mbox{d}\lambda} +1 \right) \sum_{n=0}^\infty {\rm e}^{-\lambda \beta\omega n}\nonumber\\
 &=&\frac{  1}{\left(1- \exp(-\lambda \beta\omega )\right)^{3}}.
\end{eqnarray}

\section{Color factors}\label{cf}
We shall discuss the color structure of the interaction term and the corresponding color factors here. As mentioned in the main text, the symbol $C^{aQ}$ carries the color indices of the interacting plasma constituent $a$ and both $Q$ and $\bar Q$. In the first order, the diagrams are proportional to Tr$_1 C^{aQ}$ (or Tr$_1 C^{a\bar Q}$, respectively), where $C^{aQ}$ is a product of the generators in the representations of the constituent $a$ and $Q$. For example, $C^{qQ}=t^a_{nm} t^{a\, Q}_{ji} \delta_{lk}$. Now, the effect of Tr$_1$ is that first order diagrams are proportional to the trace of the generator in the corresponding representation of the interacting constituent $a$, which vanishes. Therefore, first-order diagrams do not contribute.

\begin{figure}
\begin{center}
\begin{picture}(300,100)(0,0)
\Gluon(100,33)(200,33){5}{7}
\Gluon(100,66)(200,66){5}{7}
\ArrowLine(80,0)(100,33)
\ArrowLine(100,33)(100,66)
\ArrowLine(100,66)(80,100)
\ArrowLine(220,0)(200,33)
\ArrowLine(200,33)(200,66)
\ArrowLine(200,66)(220,100)
\Vertex(100,33){2}
\Vertex(200,33){2}
\Vertex(100,66){2}
\Vertex(200,66){2}
\Text(80,20)[r]{$m$}
\Text(80,80)[r]{$n$}
\Text(220,20)[l]{$i$}
\Text(220,80)[l]{$j$}
\Text(90,50)[r]{$o$}
\Text(210,50)[l]{$p$}
\Text(150,43)[]{$a$}
\Text(150,78)[]{$b$}
\Text(93,10)[l]{ ${\bf q}$}
\Text(207,10)[r]{ ${\bf Q}$}
\end{picture}
\end{center}
\caption{\label{qQQ} Interaction of $q$ and $Q$ in second order}
\end{figure}
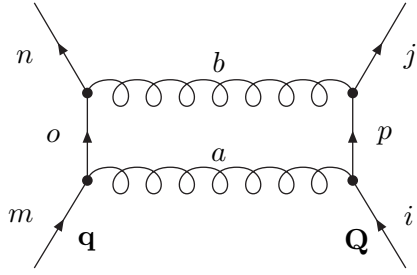

For the relevant second order, we have squares of these color structures. Let us discuss this in more detail for one example and find the corresponding color factor. Therefore we consider a quark $q$ from the plasma, interacting two times with $Q$ and assume one-gluon-exchange (see Fig. \ref{qQQ}).

The color structure is
\begin{equation}
C_{ip\,lr\,mo}^{qQ} C_{pj\,rk\,on}^{qQ}= t^a_{om} t^{a\, Q}_{pi} \delta_{lr} t^b_{no} t^{b\, Q}_{jp} \delta_{rk},
\end{equation}
where the indices $k$, $l$ and $p$ correspond to the $\bar Q$ which does not participate in this diagram. Furthermore, $t^a$ are the usual SU(3) generators in fundamental representation. With the use of $t^a_{ij} t^a_{kl}=1/2(\delta_{il}\delta_{jk}- 1/3 \delta_{ij}\delta_{lk})$ we can simplify this to
\begin{equation}
\frac12 \delta_{kl} \left( \frac59 \delta_{ij}\delta_{mn} -\frac13 \delta_{jm}\delta_{in}     \right).
\end{equation}
As we want to discuss only the $Q\bar Q$ singlet state here, we have to apply the projection operator \cite{phil}
\begin{equation}
(P_1^{Q\bar Q})_{ji, lk}=\frac19 {\delta}_{ji} {\delta}_{lk}-\frac23 t^a_{ji} (-t^a_{lk})^T.
\end{equation}
This yields the result $2/9\, \delta_{mn}$. A factor of 3, coming from the trace in Eq. (\ref{F}), is taken into account for the densities (\ref{densities}). Therefore it should not be counted twice. There are other diagrams in second order, namely interaction of $q$ with $\bar Q$, $\bar q$ with $Q$ and $\bar q$ with $\bar Q$ which give the same result of 2/9. However, there are also diagrams, where a quark $q$ interacts with both $Q$ and $\bar Q$ or an antiquark interacts with both $Q$ and $\bar Q$. These diagrams have an odd number of interactions in the ${\bf\bar 3}$ representation and therefore receive a relative minus sign.

\begin{figure}
\begin{center} 
\begin{picture}(300,100)(0,0)
\Gluon(100,33)(200,33){5}{7}
\Gluon(100,66)(200,66){5}{7}
\Gluon(80,0)(100,33){5}{3}
\ArrowLine(220,0)(200,33)
\Gluon(100,33)(100,66){5}{3}
\ArrowLine(200,33)(200,66)
\Gluon(100,66)(80,100){5}{3}
\ArrowLine(200,66)(220,100)
\Vertex(100,33){2}
\Vertex(200,33){2}
\Vertex(100,66){2}
\Vertex(200,66){2}
\Text(80,25)[r]{$a$}
\Text(80,75)[r]{$e$}
\Text(90,50)[r]{$c$}
\Text(220,25)[l]{$i$}
\Text(220,75)[l]{$j$}
\Text(210,50)[l]{$r$}
\Text(150,45)[]{$b$}
\Text(150,80)[]{$d$}
\Text(95,10)[l]{ ${\bf g}$}
\Text(205,10)[r]{ ${\bf Q}$}
\end{picture}
\end{center}
\caption{\label{gQQ} Interaction of a gluon with $Q$ in second order}
\end{figure}
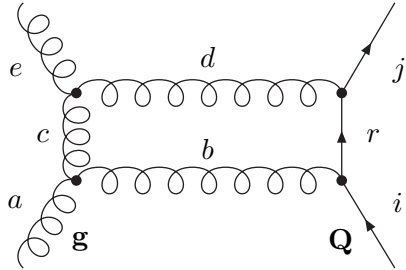

For the gluons basically the same happens. There is one diagram containing two interactions of a gluon with $Q$ and the same for $\bar Q$. In addition, there are two diagrams for interactions both with $Q$ and $\bar Q$ which have a relative minus sign for the same reasons as above. As gluons are in adjoint representation, we have to use the corresponding generators. Let us again consider an example, where a gluon interacts twice with $Q$ (see Fig. \ref{gQQ}).
The color structure reads
\begin{equation}
C^{gQ}_{ir\,kp\,ac} C^{gQ}_{rj\,pl\,ce}=t^{b\,Q}_{ri}\delta_{kp} i f^{bac} t^{d\, Q}_{jr}\delta_{pl} i f^{dce}.
\end{equation}
With the immediate use of the projection operator for the $Q\bar Q$ singlet state and the Casimir operator in adjoint representation $f^{abc} f^{ebc}=3 \delta^{ae}$, one finally gets $1/2 \delta_{ae}$. Again, the trace over color indices of the gluon was taken into account for the densities and should not be counted twice.

Diagrams which contain interactions of two plasma constituents do not contribute in second order because they are products of first-order diagrams which vanish.

\section{Calculation of the second order $F_2$}\label{app:F2}
We now calculate the relevant second order for the free energy
\begin{equation}
F_2=-{1\over 2T} \langle {\cal V}^2({\bf r}_1,\ldots,{\bf r}_N)\rangle.
\end{equation}
As we discussed before, only terms in the square which come from one plasma constituent $a$ contribute. This yields
\begin{eqnarray}\label{V2}
\hspace{-0.5cm}\langle {\cal V}^2({\bf r}_1, \dots, {\bf r}_N)\rangle &=& \frac1V  \sum_{i=1}^N c_i  \int d^3 {\bf r}_i 
\left[ 2 \alpha_s^2 \left(\frac{
\exp(-m_D \vert{\bf r}_i- {\bf x}_Q\vert)}{\vert{\bf r}_i- {\bf x}_Q\vert}\right)^2\nonumber\right.\\
&&\left. - 2 \alpha_s \frac{\exp(-m_D \vert{\bf r}_i- {\bf x}_{ Q}\vert)}{\vert{\bf r}_i- {\bf x}_Q\vert} \alpha_s\frac{\exp(-m_D \vert{\bf r}_i- {\bf x}_{\bar Q}\vert)}{\vert{\bf r}_i- {\bf x}_{\bar Q}\vert}                                                                 \right].
\end{eqnarray}
Here, $V$ is a three-dimensional volume (coming from $F_0$) which will be absorbed in the densities. While the first term comes from the Yukawa potential squared, the second term corresponds to the mixed term of the square. With the use of Fourier transformation, we can proceed to (${\bf d}_i={\bf r}-{\bf x}_i$)
\begin{eqnarray}
&&-2 (4\pi)^2 \int d^3 {\bf r}_i {d^3 {\bf q}_1\over (2\pi)^3} {d^3 {\bf q}_2\over (2\pi)^3} \alpha_s(q_1, T)\frac{\exp(i{\bf q}_1{\bf d}_Q)}{(q_1^2+m_D^2)^2} \alpha_s(q_2, T)\frac{\exp(i{\bf q}_2{\bf d}_{\bar Q})}{(q_2^2+m_D^2)^2} \nonumber\\
&=&- \frac4\pi \int d^3{\bf q} \alpha_s(q, T)^2 \frac{\exp(i {\bf q} ({\bf x}_Q -{\bf x}_{\bar Q}))}{(q^2+m_D^2)^2}
\end{eqnarray}
for the second term and
\begin{eqnarray}
&&2 \int  d^3{\bf r}_i \left(\alpha_s( \vert{\bf r}_i- {\bf x}_Q\vert ,T)\frac{
\exp(-m_D \vert{\bf r}_i- {\bf x}_Q\vert)}{\vert{\bf r}_i- {\bf x}_Q\vert}\right)^2\nonumber\\
&=& \frac4\pi \int d^3{\bf q} \frac{\alpha_s(q, T)^2 }{(q^2+m_D^2)^2}
\end{eqnarray}
for the first one. The sum becomes trivial for each species of plasma constituents and can be written as $\sum_a c_a N_a^{{\rm eff}}$. With $N_a^{{\rm eff}}/V=n_a^{{\rm eff}}$ one can then introduce the densities here.
 Altogether we have
\begin{eqnarray}
F_2&=&-{1\over 2T} \langle V^2({\bf r}_1, \dots , {\bf r}_N)\rangle\nonumber\\
&=& -\frac{2}{\pi T} \sum_a c_a n_a^{{\rm eff}} \int d^3 {\bf q} \frac{\alpha_s(q, T)^2 }{(q^2+m_D^2)^2}\left(1- {\rm e}^{i {\bf q}( {\bf x}_Q- {\bf x}_{\bar Q})}   \right).
\end{eqnarray}

\section{Computation of the density of free relativistic particles}\label{ns}
We want to calculate the density of free relativistic particles for  quarks (fermions) and gluons (bosons).
It has the form 
\begin{equation}\label{nn}
n_a=D_a\int {d^3 {\bf p}\over (2\pi)^3} \frac{1}{\exp(\beta E_0 ) \pm 1}
\end{equation}
where $a$ stands for $q$, $\bar q$ or $g$, $E_0=\sqrt{{\bf p}^2 +m_a^2}$, and $D_a$ is the degeneracy factor. Explicitly, this factor is $(N_c^2-1) N_p =8 \cdot 3=24$ (due to color and  polarization) for (massive) gluons and $ N_c N_f (2s_q+1)=  3\cdot N_f \cdot 2=6 N_f$ (due to  color, flavors and spin) for quarks. 

Let us begin with quarks.
The denominator can be rewritten by means of the geometric series:
\begin{equation}
\frac{1}{\exp(\beta E_0 ) + 1}=\frac{1}{\exp(\beta E_0 )}\frac{1}{1+\exp(-\beta E_0 )}
= {\rm e}^{-\beta E_0 } \sum_{n=0}^\infty (-1)^n {\rm e}^{-\beta n E_0}. 
\end{equation}
Next, we can use the general formula 
\begin{equation}\label{commonint}
\int_0^\infty\frac{da}{\sqrt{\pi a}}{\rm e}^{-m^2 a-\frac{R^2}{4a}}=\frac{{\rm e}^{-m R}}{m}
\end{equation} 
to obtain
\begin{eqnarray}
&n_q&=12 N_f\sum_{n=1}^\infty (-1)^{n+1}\int {d^3 {\bf p}\over (2\pi)^3}  {\rm e}^{-\beta n E_0} \nonumber\\
	&&=12 N_f  \sum_{n=1}^\infty (-1)^{n+1}\int {d^3 {\bf p}\over (2\pi)^3} \int_0^\infty
	\frac{da}{\sqrt{\pi a}}\exp\left(- a-\frac{\beta^2 n^2 ({\bf p}^2 +m_q^2)}{4a}\right).
\end{eqnarray}
The integral over ${\bf p}$ is now Gaussian and gives $\frac{1}{(2\pi)^3}\left(\frac{2\sqrt{\pi a}}{\beta n}\right)^3$. With the use of the formula
\begin{equation}\label{bessel}
\int_0^\infty dt t^{\alpha-1} \exp\left(-\gamma t- {\delta \over t}\right)=2 \left(\frac{\delta}{\gamma}\right)^{\alpha/2} K_\alpha(2 \sqrt{\gamma\delta}),~~ {\rm Re}{\,}\gamma>0,~~ {\rm Re}{\,}\delta>0,
\end{equation}
we get the final result
\begin{equation}
n_q=3 N_f T \frac{m_q^2}{\pi^2}  \sum_{n=1}^\infty \frac{(-1)^{n+1}}{n} K_2\left({m_q\over T} n \right).
\end{equation}
The same result will appear for antiquarks, with $m_{\bar q}$ instead of $m_q$.
For gluons (bosons) with the thermal mass $m_g$ the computation is nearly the same. Due to the minus sign in the denominator of (\ref{nn}), the factor $(-1)^n$ does not appear (and, of course, the factor of degeneracy is changed). We have
\begin{equation}
n_g=12 T \frac{m_g^2}{\pi^2}  \sum_{n=1}^\infty \frac{1}{n} K_2\left({m_g\over T} n \right).
\end{equation}

\end{appendix}


\begin{thebibliography}{999}

\bibitem{Adcox:2004mh}
  K.~Adcox {\it et al.}  [PHENIX Collaboration],
  Nucl.\ Phys.\ A {\bf 757}, 184 (2005) 

\bibitem{Arsene:2004fa}
  I.~Arsene {\it et al.}  [BRAHMS Collaboration],
  Nucl.\ Phys.\ A {\bf 757}, 1 (2005) 


\bibitem{Back:2004je}
  B.~B.~Back {\it et al.} [PHOBOS Collaboration],
  Nucl.\ Phys.\ A {\bf 757}, 28 (2005) 

\bibitem{Adams:2005dq}
  J.~Adams {\it et al.}  [STAR Collaboration],
  Nucl.\ Phys.\ A {\bf 757}, 102 (2005) 

\bibitem{revs} 
For a recent review see: E.~V.~Shuryak, {\it The QCD
    vacuum, hadrons and superdense matter. 2nd edition}, World
  Scientific (2004)

\bibitem{ay}
P.~Arnold, G.~D.~Moore and L.~G.~Yaffe,
JHEP {\bf 11}, 001 (2000); ibid. {\bf 05}, 051 (2003)

\bibitem{yuasim}
Yu.~A.~Simonov,
Phys.\ Lett.\ B {\bf 619}, 293 (2005)


\bibitem{latint1}
P.~Petreczky and K.~Petrov,
Phys.\ Rev.\ D {\bf 70}, 054503 (2004)

\bibitem{lat}
P.~Petreczky,
Eur.\ Phys.\ J.\ C {\bf 43}, 51 (2005)


\bibitem{L}
K.~J.~Juge, J.~Kuti and C.~Morningstar, 
Nucl.\ Phys.\ Proc.\ Suppl.\  {\bf 63}, 326 (1998);  ibid. {\bf 73}, 590 (1999);
Phys.\ Rev.\ Lett.\  {\bf 82}, 4400 (1999);
 ibid.  {\bf 90}, 161601 (2003)

\bibitem{spect}
N.~Brambilla, A.~Pineda, J.~Soto and A.~Vairo,
Nucl.\ Phys.\ B {\bf 566}, 275 (2000);
Rev.\ Mod.\ Phys.\  {\bf 77}, 1423 (2005);
Yu.~S.~Kalashnikova and D.~S.~Kuzmenko,
Phys.\ Atom.\ Nucl.\  {\bf 64}, 1716 (2001); 
ibid. {\bf 66}, 955 (2003)

\bibitem{Greensite}
J.~Greensite and C.~B.~Thorn, JHEP {\bf 02}, 014 (2002)


\bibitem{idr}
C. Itzykson and J.-M. Drouffe, {\it Statistical field theory. Vol. 1}, Cambridge Univ. Press (1989)


\bibitem{corrlength}
A.~Di~Giacomo and H.~Panagopoulos, Phys.\ Lett.\ B {\bf 285}, 133 (1992);
M.~D'Elia, A.~Di Giacomo and E.~Meggiolaro,
Phys.\ Lett.\ B {\bf 408}, 315 (1997);
for a review see, e.g., A.~Di~Giacomo, "Nonperturbative QCD" [arXiv:hep-lat/9912016]; 
for the recent finite-temperature measurements see:  M.~D'Elia, A.~Di Giacomo and E.~Meggiolaro,
Phys.\ Rev.\ D {\bf 67}, 114504 (2003)

\bibitem{pa}
R.~D.~Pisarski and O.~Alvarez,
Phys.\ Rev.\  D {\bf 26} 26,  3735  (1982)


\bibitem{MO}
See, e.g., H.~Meyer-Ortmanns,
Rev.\ Mod.\ Phys.\  {\bf 68}, 473 (1996)

\bibitem{DFP}
S.~Digal, S.~Fortunato and P.~Petreczky,
Phys.\ Rev.\  D {\bf 68}, 034008 (2003)

\bibitem{CDM}
M.~Campostrini, A.~Di Giacomo and G.~Mussardo,
Z.\ Phys.\  C {\bf 25}, 173 (1984)

\bibitem{tc}
O.~Kaczmarek and F.~Zantow, Eur.\ Phys.\ J.\ C {\bf 43}, 63 (2005);
"Static quark anti-quark interactions at zero and finite temperature QCD.
II: Quark anti-quark internal energy and entropy"
[arXiv:hep-lat/0506019]
  
\bibitem{const1}
D.~Gromes,
"Theoretical understanding of quark forces", 
preprint HD-THEP-89-17

\bibitem{const2}
H.~J.~Pirner and M.~Wachs, Nucl.\ Phys.\ A {\bf 617}, 395 (1997) [arXiv:hep-ph/9701281]
  

\bibitem{bpir}
J.~Braun and H.~J.~Pirner,
"Energy loss in the Quark-Gluon Plasma" [arXiv:hep-ph/0610331]

\bibitem{bgies}
J.~Braun and H.~Gies, JHEP {\bf 06}, 024, (2006)

\bibitem{landau}
 L.~D.~Landau and E.~M.~Lifshitz, {\it Statistical physics}, Pergamon (1958)


\bibitem{lebellac}
M.~Le Bellac, {\it Thermal field theory}, Cambridge Univ. Press (1996)

\bibitem{peshier}
A.~Peshier, B.~K\"ampfer, G.~Soff, 
Phys.~Rev.~C {\bf 61}, 045203 (2000)

\bibitem{karsch}
F.~Karsch, E.~Laermann and A.~Peikert,
Phys.\ Lett.\ B {\bf 478}, 447 (2000) 



\bibitem{shu:bound}
E.V.~Shuryak, I.~Zahed, Phys.~Rev.~D {\bf 70}, 054507 (2004) 

\bibitem{k:bound}
S.~Datta, F.~Karsch, P.~Petreczky, I.~Wetzorke, Phys.~Rev.~D {\bf 69}, 094507 (2004); 
F.~Karsch, Eur.~Phys.~J. C {\bf 43}, 35 (2005) 

\bibitem{c:bound}
N. Brambilla et al., "Heavy quarkonium physics" [arXiv:hep-ph/0412158]


\bibitem{phil}
O.~Philipsen, Phys.~Lett.~B {\bf 535}, 138 (2002)

\end{thebibliography}
\end{document}